\documentclass[sn-mathphys]{sn-jnl}

\usepackage{graphicx}
\usepackage{booktabs}
\usepackage{multirow}
\usepackage{xcolor}
\usepackage{soul}

\begin{document}

\title[Understanding Policy and Technical Aspects of AI-Enabled SVS]{Understanding Policy and Technical Aspects of AI-Enabled Smart Video Surveillance to Address Public Safety}

 \author*[1]{\fnm{Babak} \sur{Rahimi~Ardabili}}\email{brahimia@uncc.edu}

 \author[2]{\fnm{Armin} \sur{Danesh~Pazho}}\email{adaneshp@uncc.edu}

 \author[2]{\fnm{Ghazal} \sur{Alinezhad~Noghre}}\email{galinezh@uncc.edu}

 \author[2]{\fnm{Christopher} \sur{Neff}}\email{cneff1@uncc.edu}

 \author[2]{\fnm{Sai} \sur{Datta Bhaskararayuni}}\email{sbhaska1@uncc.edu}

 \author[2]{\fnm{Arun} \sur{Ravindran}}\email{arun.ravindran@uncc.edu}

 \author[3]{\fnm{Shannon} \sur{Reid}}\email{s.reid@uncc.edu}

 \author[2]{\fnm{Hamed} \sur{Tabkhi}}\email{htabkhiv@uncc.edu}

 \affil*[1]{\orgdiv{Public Policy Program}, \orgname{University of North Carolina at Charlotte}, \orgaddress{\street{9201 University City Blvd}, \city{Charlotte}, \postcode{28223}, \state{North Carolina}, \country{US}}}

 \affil[2]{\orgdiv{Electrical Engineering and Computer Systems}, \orgname{University of North Carolina at Charlotte}, \orgaddress{\street{9201 University City Blvd}, \city{Charlotte}, \postcode{28223}, \state{North Carolina}, \country{US}}}

 \affil[3]{\orgdiv{Criminal Justice}, \orgname{University of North Carolina at Charlotte}, \orgaddress{\street{9201 University City Blvd}, \city{Charlotte}, \postcode{28223}, \state{North Carolina}, \country{US}}}


\abstract{Recent advancements in artificial intelligence (AI) have seen the emergence of smart video surveillance (SVS) in many practical applications, particularly for building safer and more secure communities in our urban environments. Cognitive tasks, such as identifying objects, recognizing actions, and detecting anomalous behaviors, can produce data capable of providing valuable insights to the community through statistical and analytical tools. However, artificially intelligent surveillance systems design requires special considerations for ethical challenges and concerns. The use and storage of personally identifiable information (PII) commonly pose an increased risk to personal privacy. To address these issues, this paper identifies the privacy concerns and requirements needed to address when designing AI-enabled smart video surveillance.
Further, we propose the first end-to-end AI-enabled privacy-preserving smart video surveillance system that holistically combines computer vision analytics, statistical data analytics, cloud-native services, and end-user applications. Finally, we propose quantitative and qualitative metrics to evaluate intelligent video surveillance systems. The system shows the 17.8 frame-per-second (FPS) processing in extreme video scenes. However, considering privacy in designing such a system results in preferring the pose-based algorithm to the pixel-based one. This choice resulted in dropping accuracy in both action and anomaly detection tasks. The results drop from 97.48 to 73.72 in anomaly detection and 96 to 83.07 in the action detection task. On average, the latency of the end-to-end system is 36.1 seconds.  
}


\maketitle

\section{Introduction}\label{sec1}

The emergence of new technologies and developments in implementing these technologies affected different aspects of our life\cite{aslania2016modeling}. The emergence of new concepts such as digital health, smart transportation, and smart city are examples of these effects. These new technologies make the current systems more efficient and provide more opportunities in each ecosystem. For example, using AI in healthcare systems provides new and more efficient healthcare solutions in diagnosing cancer diseases \cite{huang2020artificial}. Although many of these technologies and new trends are different from a technical perspective, they are similar in their dependence on data. They are dependent on data as input, and they also generate valuable information by digesting the input data, which could be used as inputs in other systems. 

\begin{figure*}[h]
        \centering
               \includegraphics[width=1\linewidth]{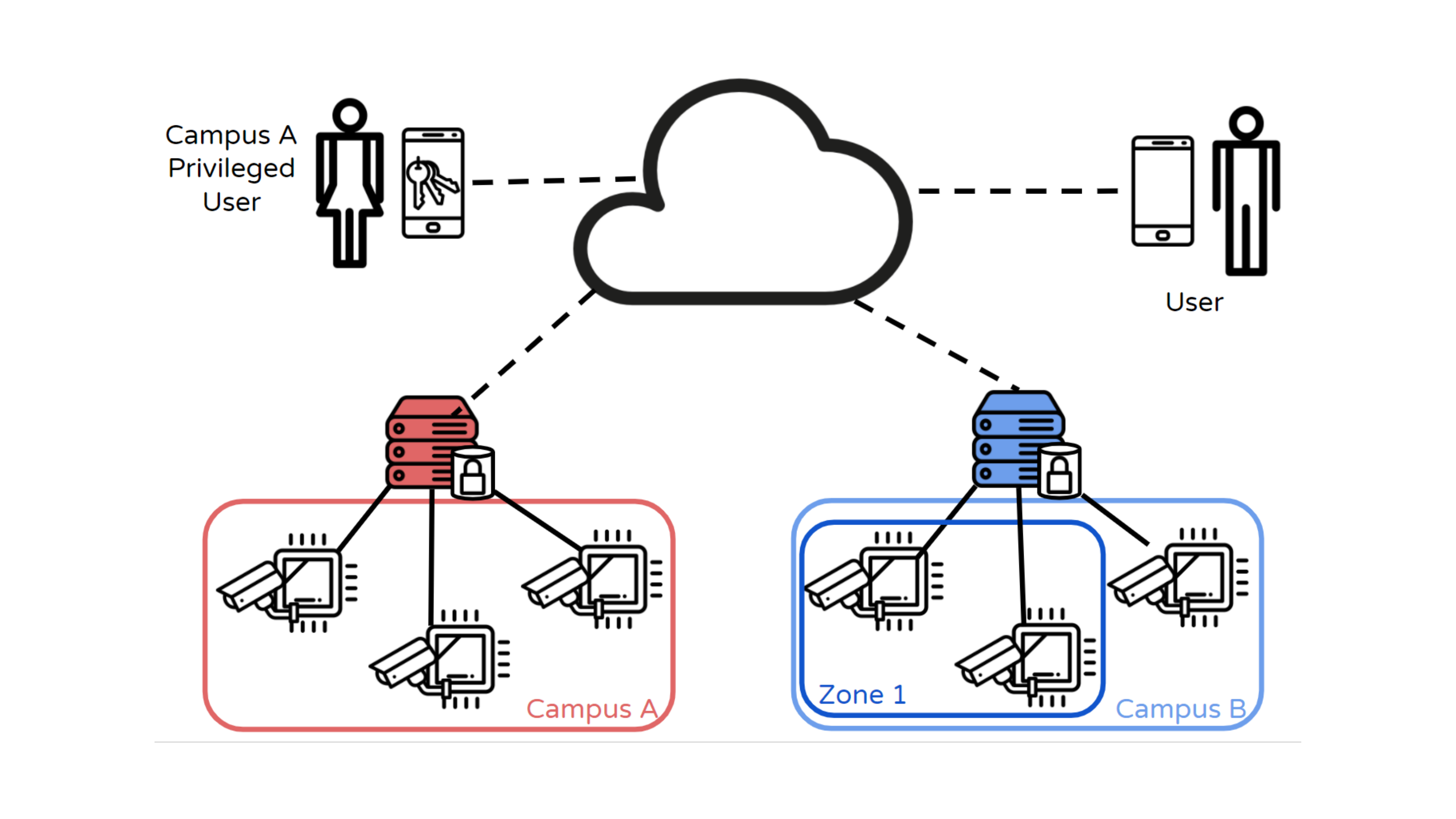}
                \caption{Smart Video Surveillance General Structure}
                \label{fig:SVS}
\end{figure*}

Adopting new technologies in the smart city sector has altered some city functions and developed new ones. From the public point of view, creating a safe community as an essential city function is one of the most crucial goals of city officials \cite{fraser2018goals}. Increasing policing and monitoring citizens' daily activities is the classic approach toward public safety. Traditionally in public spaces requiring monitoring, installing Closed Circuit TVs (CCTV) and having people manually check the camera feeds are the most common approach. However, with recent advances in AI, new opportunities have been developed to achieve this city goal. SVS technology is developed to help law enforcement by increasing situational awareness \cite{ardabili2022understanding}.

SVS systems not only help make monitoring processes more efficient but also provide opportunities to collect different types of data \cite{cangialosi2022privid} that are crucial in making a safer community. Different organizations and agencies need various types of data to play their roles in improving public safety. Detecting anomalies, recognizing actions, detecting objects, identifying criminal behaviors, geo-location data of individuals as well as anomalous events, and real-time occupancy are examples of data that is provided through SVS technologies\cite{ardabili2022understanding,zhang2014context,cangialosi2022privid}. The potential of SVS technology in collecting and generating the required data resulted in increasing global attention to developing and using this technology. According to the market predictions reports, the global video surveillance market expects to reach 144.85 billion USD value by 2027, experiencing the 14.6 percent Compound Annual Growth Rate (CAGR) \cite{fathy2022integrating}. 

Providing the pre-mentioned data through implementing SVS technology increases the risk of privacy violation. The privacy risk can be understood from two perspectives: the type of input data and the organizations that use the data \cite{cangialosi2022privid}. In the surveillance context, cameras record the pedestrians' and citizens' images and frames to collect the data. This process inherently increases the risk of storing personal images and tracking citizens' activities \cite{nissenbaum2004privacy}. On the other hand, many organizations need access to camera footage to fulfill their goals. Security firms, transportation agencies \footnote{https://www.govtech.com/fs/data/video-analytics-traffic-study-creates-baseline-for-change.html}, and shopping malls and grocery stores \footnote{https://www.briefcam.com/resources/blog/how-retail-stores-can-streamline-operations-with-video-content-analytics/} are examples of these industries. Either the firms misuse the shared video records \cite{martin2017data}or some staff in the firm illegally use them \cite{moore2015creepy}, individuals' privacy is violated. Therefore, there is a long-lasting debate between privacy concerns and SVS stakeholders to prevent privacy violations at both public and private entities \cite{hartzog2018privacyos}. 

Scholars, businesses, and policymakers consider privacy concerns through different strategies to ensure citizens and users that their privacy concerns are being addressed at different levels \cite{leenes2019regulating, almeida2022ethics,nissenbaum2004privacy,acquisti2015privacy}. At the policy level, although there is a lack of federal regulation to address the issue in the SVS context, some privacy protection policies have been issued at the state level as a response to this concern. The use of facial recognition technology, for instance, is banned or limited in law enforcement in the states of California, Vermont, Virginia, Massachusetts, Maine, New York, Washington, Maryland, and Oregon. As a result of these regulations and policies, some big technology companies such as Facebook, Microsoft, and IBM stopped or limited the proposing and offering of tools and solutions that use facial recognition technologies \cite{almeida2022ethics}. On the other hand, scholars and businesses developed different tools and algorithms, such as face blurring tools, to limit using personal information in SVS \cite{padilla2015visual, wu2021pecam}.

Considering privacy at the system design stage is one of the practical approaches to addressing the privacy challenges in the SVS context. Incorporating privacy issues at this stage result in more flexible solutions to solve the problem\cite{hartzog2018privacyos}. Incorporating privacy concerns in the design phase requires building a system from scratch and selecting every detail of the system design with privacy protection consideration. In this paper, we propose an end-to-end SVS system design. The system is designed to use pre-installed cameras in public places such as parking lots to collect the required data for improving public safety. Figure \ref{fig:SVS} shows the general structure of such system. This system uses different elements and algorithms to deliver the data to the end user. Identifying privacy needs and requirements to materialize AI-enabled smart video surveillance is our first contribution to this paper. Based on that, we propose the first end-to-end AI-enabled privacy persevering smart video surveillance system that holistically combines computer vision analytics, statistical data analytics, cloud-native services, and smartphone application as essential elements in helping the public make a safer community. Finally, we are not only proposing a prototype of such a system, but also we propose the quantitative and qualitative metrics to evaluate this system. Our main contributions in this paper can be summarized as: 
Lists are easy to create:
\begin{itemize}
  \item Identifying privacy needs and requirements; 
  \item Proposing the first end-to-end AI-enabled privacy persevering smart video surveillance system;
  \item Proposing the quantitative and qualitative metrics to evaluate the system.
\end{itemize}

\section{Related Works}\label{sec14}

\subsection{Academic Background}\label{subsec2}
For many years, SVS has been a popular research topic. Researchers from the University of Alcal\'a \cite{MallBehaviors} proposed a system for real-time detection of suspicious behaviors in shopping malls in 2015, employing a combination of artificial intelligence approaches to track individuals through a mall's security system and determine when suspicious behaviors occur. However, as with many other works in the field, no consideration is given to ethical concerns, the privacy of those being tracked, or any bias that the system may learn. Instead, the research focuses on real-time execution and achieving high accuracy on publicly available datasets. Peeking into the Future \cite{Peeking} recently proposed an end-to-end system for predicting people's actions in a video surveillance setting. While the work does not address ethical, privacy, or fairness concerns, the authors conclude that future work geared toward real-world applications may need to prioritize these concerns.

The emphasis in REVAMP\textsuperscript{2}T \cite{REVAMP2T} is on performing person re-identification and tracking in a multi-camera environment while protecting the privacy of the people being tracked. To accomplish this, they propose two policies. The first is that no image data is stored or transferred across the network; image data is destroyed once the system processes it. They claim that this prevents even those with direct access to the system from accessing a person's personally identifiable information. The second difference is that instead of using invasive technologies to identify individuals (such as facial recognition), their re-identification algorithm employs an abstract representation of a person's features that humans cannot interpret.
In this way, their goal was to emphasize differentiation between people rather than personal identification. Other works have taken similar approaches and directly applied them to the field of SVS \cite{TX2MTMC}.

In some works, authors proposed privacy-preserving systems. However, their approach is different from our proposed system in that they did not incorporate privacy protection in their system design process. For example, Privid is a proposed privacy-perseverance system design \cite{cangialosi2022privid}. However, the authors did not incorporate the privacy-by-design approach to their work. In this approach, they calculated the time and location in the frame occupied by people more and masked that specific part of the scene rather than masking and blurring people. As another example, Gupta and Prabhat (2022) proposed a privacy-preserving system\cite{gupta2022towards}. Their contribution focuses on optimizing the resources and server side of the system.  They also obfuscated individual faces as their privacy-preserving strategy. 

\subsection{Industrial Background}\label{subsec3}
SVS systems have been widely used in a variety of industries. Although these systems have some industrial applications, addressing privacy concerns is not the primary goal of their proposed systems. The majority of companies in the SVS sector provide various video management services. As a security solution, they typically provide a built-in feature integrated into their general service.

Some companies provide services to blur the actual videos. For example, Milestone systems include a "privacy masking" feature to protect privacy. This function includes a modular blurring algorithm. The user can choose whether or not to blur the video and the intensity of the mask\footnote{https://www.milestonesys.com/}. The actual videos are still accessible in this configuration.

A few businesses offer SVS-based crime detection solutions. They provide services such as object and person detection and action recognition. They rarely provide information about the data and algorithms they use \footnote{https://getsafeandsound.com/2021/01/top-video-surveillance-companies-2021/}. Avigilion Corporation is one of these companies. Avigilion has a search feature that allows customers to look for specific people and license plates in images\footnote{https://www.avigilon.com/}. Misuse of this feature may infringe on an individual's privacy rights.

Some businesses provide privacy perseverance systems but must provide detailed information about their approach to designing a privacy perseverance system. For example, Genetec Omnicast provides video surveillance management services to clients. As a feature of the service, Genetec mentions privacy and security protection \footnote{https://www.genetec.com/}. However, they primarily discussed the products' cyber-security features.

\section{Privacy and Regulations}\label{PR}
It is worth noting that no federal law has yet to address privacy issues from a technical standpoint. Some regulations, however, have been developed to assist developers in ensuring that the technology complies with public privacy concerns. The most important acts in the United States that address this issue in various sectors are the Health Insurance Portability and Accountability Act (HIPAA), the California Consumer Privacy Act (CPPA), and the American Data Privacy and Protection Act (ADPPA). The General Data Protection Regulation (GDPR), the European Union's set of data privacy and protection rules, is, on the other hand, the most important act in Europe.

The HIPAA Privacy Rule is a national standard that safeguards medical records and other personally identifiable health information. HIPAA applies to all providers who use electronic healthcare transactions, including health plans and healthcare provider centers. The goal of HIPAA is to establish specific rules to protect individuals' privacy. In general, HIPAA establishes standards to protect individuals' rights to prohibit the unauthorized use of health information and to access their medical and health records to obtain a copy, request corrections, and transmit the electronic version to a third party \cite{hipaa}. 

The California Consumer Privacy Act, enacted in 2018, was the first comprehensive commercial privacy law. The CPPA's goal is to provide clear guidelines for organizations and consumers in California. It will apply to any for-profit entity in California that collects, shares, or sells personal data from California consumers and has annual gross revenues of more than 25 million USD. The CCPA is silent about the Protected Health Information (PHI) collected by covered entities or business associates, leaving it subject to HIPAA. It also excludes medical information covered by California's equivalent law, the Confidentiality of Medical Information Act (CMIA). The California Consumer Privacy Rights Act (CPRA) was passed in 2020. CPRA expands CPPA by allowing consumers to
Request that businesses do not share their personal information;
Request to correct the incorrect personal information; and
Restrict enterprises' use of "sensitive personal information," which includes geolocation, race, ethnicity, religion, genetic data, private communications, sexual orientation, and specified health information.

The American Data Privacy and Protection Act (ADPPA) has yet to be passed by Congress, but it is expected to go into effect soon. As a result, they are investigating the ADPPA's perspective on privacy as the most recent act is critical. Most entities, including nonprofits, common carriers, large data holders, and service providers, would be covered by the bill. The ADPPA would govern how businesses store and use consumer data. This act requires data collectors to limit the amount of data they collect unless it is "necessary, proportionate, and limited to" their business purpose.
ADPPA restricts the transfer and, in some cases, processing of Social Security numbers, precise geolocation, biometric and genetic data, passwords, browsing history, and physical activity tracking \footnote{All content is derived from https://www.congress.gov/}.

The General Data Protection Regulation (GDPR) is a European Union (EU) data and privacy protection regulation that applies to the EU and the European Economic Area (EEA). The GDPR's goal is to establish standards that ensure individuals have control over their data while simplifying the international business regulatory environment. The GDPR's rules apply to all "personal data." Personal data is "any information relating to a living, identified, or identifiable person" under GDPR. Personal data includes the subject's name, Social Security Number (SSN), other identification numbers, location data, IP addresses, online cookies, images, email addresses, and content generated by the data subject\cite{viorescu20172018}. 

As we can see, these regulations fall outside the SVS scope. However, they provide a reasonable starting point for evaluating and addressing privacy issues in SVS.

None of the regulations explicitly discussed address the privacy issue in the context of SVS. However, summarizing existing regulations and policy perspectives provides a framework for addressing the context's privacy concerns. According to the preceding section, privacy can be addressed from the perspectives of the algorithm, system, model, and data.

As shown in table \ref{tab:sum}, all acts prohibiting entities from using identifiable information are prohibited. As a result, from an algorithm standpoint, the best algorithms do not rely on identifiable information. The system should be designed to prevent data from being transferred to a third party. This system complies with all applicable laws. These acts also mention data retention and data irreversibility. These two issues should be addressed by the models that are used. Finally, the data type is critical. They should be de-identified information. As a result, personally identifiable information and facial recognition technology should not be used in the design of a compliance system.

\begin{table*}[]
\centering
\caption{Summary of privacy protection acts (GDPR\cite{viorescu20172018}, HIPAA\cite{hipaa}, CCPA\cite{10.2307/j.ctvjghvnn}, ADPPA is derived from https://www.congress.gov/.}
\label{tab:sum}
\resizebox{\textwidth}{!}{%
\begin{tabular}{c|c|c|c|c|c|c}
\textbf{Regulations} & \textbf{Domain}                                                & \textbf{Type of Data}                                     & \textbf{Data Collection}                                              & \textbf{Data Transfer}                                                           & \textbf{Data Processing}                                              & \textbf{Data Retention}                                                           \\ \hline \hline
GDPR                 & General                                                        & Personal data                                             & Minimized                                                             & \begin{tabular}[c]{@{}c@{}}By user's \\ contest\end{tabular}                     & \begin{tabular}[c]{@{}c@{}}Consistent with\\ the purpose\end{tabular} & \begin{tabular}[c]{@{}c@{}}Consistent with\\ the purpose\end{tabular}             \\ \hline
HIPAA                & \begin{tabular}[c]{@{}c@{}}Health \& \\ Insurance\end{tabular} & \begin{tabular}[c]{@{}c@{}}Medical\\ records\end{tabular} & \begin{tabular}[c]{@{}c@{}}Consistent with\\ the purpose\end{tabular} & \begin{tabular}[c]{@{}c@{}}Allowed \\ between authorized\\ entities\end{tabular} & \begin{tabular}[c]{@{}c@{}}Allowed\\ by covered entities\end{tabular} & \begin{tabular}[c]{@{}c@{}}No fewer \\ than six years\end{tabular}                \\ \hline
CCPA                 & \begin{tabular}[c]{@{}c@{}}General\\ CL residents\end{tabular} & Personal data                                             & \begin{tabular}[c]{@{}c@{}}By informing\\ the consumers'\end{tabular} & \begin{tabular}[c]{@{}c@{}}Allowed by\\ prior notice\end{tabular}                & \begin{tabular}[c]{@{}c@{}}By \\ pseudonymization\end{tabular}        & \begin{tabular}[c]{@{}c@{}}By consumer's\\ request\end{tabular}                   \\ \hline
ADPPA  & General    & Personal data       & Minimized                                                             & \begin{tabular}[c]{@{}c@{}}Deidentified\\ data are \\ allowed\end{tabular}       & \begin{tabular}[c]{@{}c@{}}Consistent with\\ the purpose\end{tabular} & \begin{tabular}[c]{@{}c@{}}At the end \\ of the service \\ or by law\end{tabular} \\ 
\end{tabular}
}
\end{table*}

\section{Proposed Privacy-Preserving Smart Video Surveillance System}\label{sec15}
\begin{figure*}[h]
        \centering
               \includegraphics[width=1\linewidth]{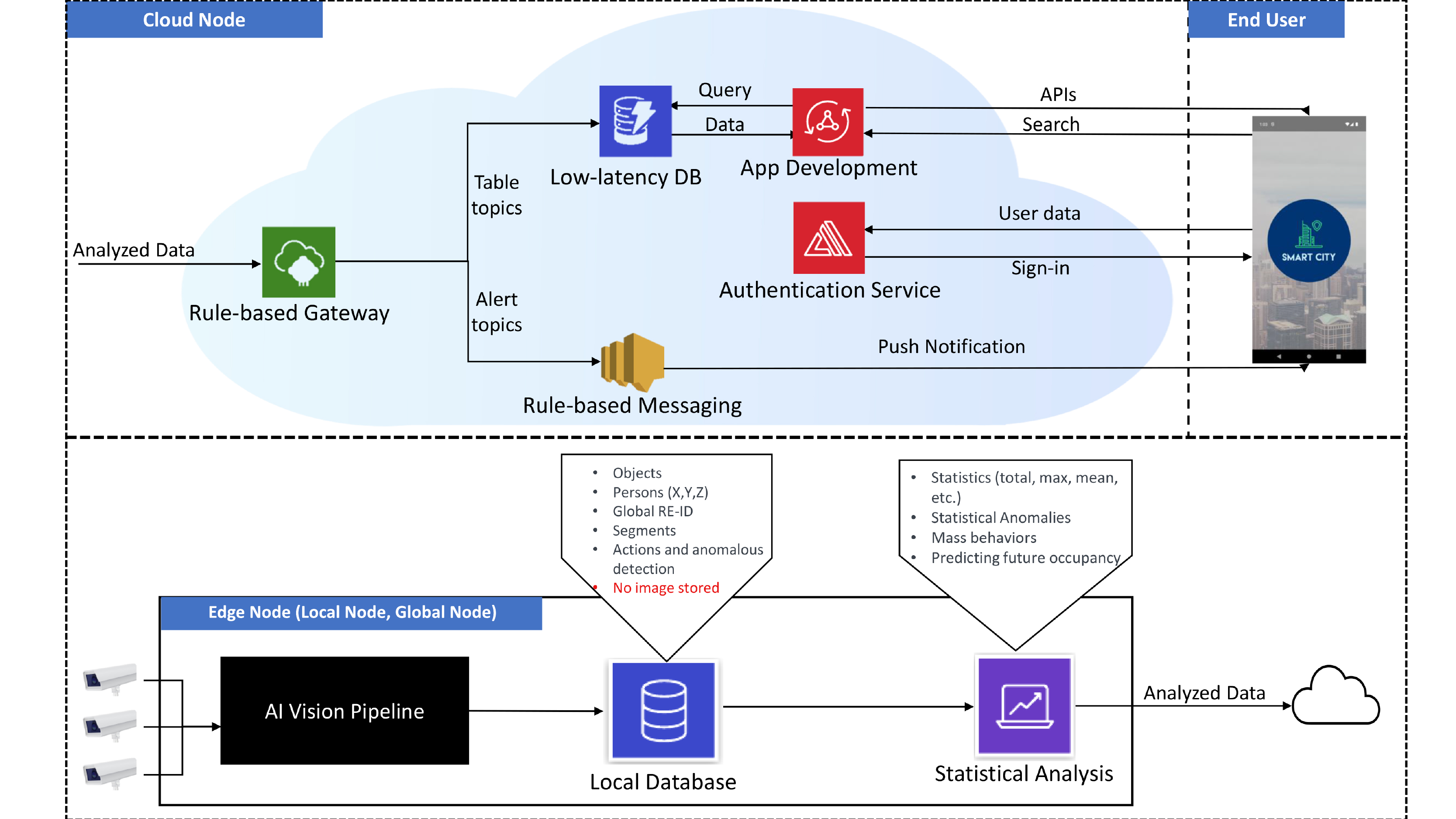}
                \caption{The Proposed End-to-End System Design}
                \label{fig:E2E}
\end{figure*}

In this section, we outline our proposed end-to-end SVS system. As we discussed in the introduction, the main feature of the proposed system is using the footage of pre-existing cameras in public places to generate required data that could be used to improve public safety. As discussed earlier in this paper, we need a system that generates data such as detected actions, anomalous behaviors, detected objects, and the number of tracked people. The cameras use real-time images of pedestrians. Therefore, the algorithms, services, and system features are selected to protect privacy.

This system consists of three main sections, as represented in figure \ref{fig:E2E}. The first section is the edge node. The AI pipeline, edge database, and statistical models are hosted on the edge server. The results of the analysis will be sent to the cloud node. In the cloud node, different cloud services host the smartphone application. Using cloud-native services is a cost-effective and scalable strategy for hosting smartphone applications. Finally, the smartphone application delivers the required data to the end users. In the next sections, we discuss the detailed information in each section.

\subsection{Edge Node}\label{edgenode}
\begin{figure*}[h]
        \centering
               \includegraphics[width=1\linewidth, trim= 26 18 18 24,clip]{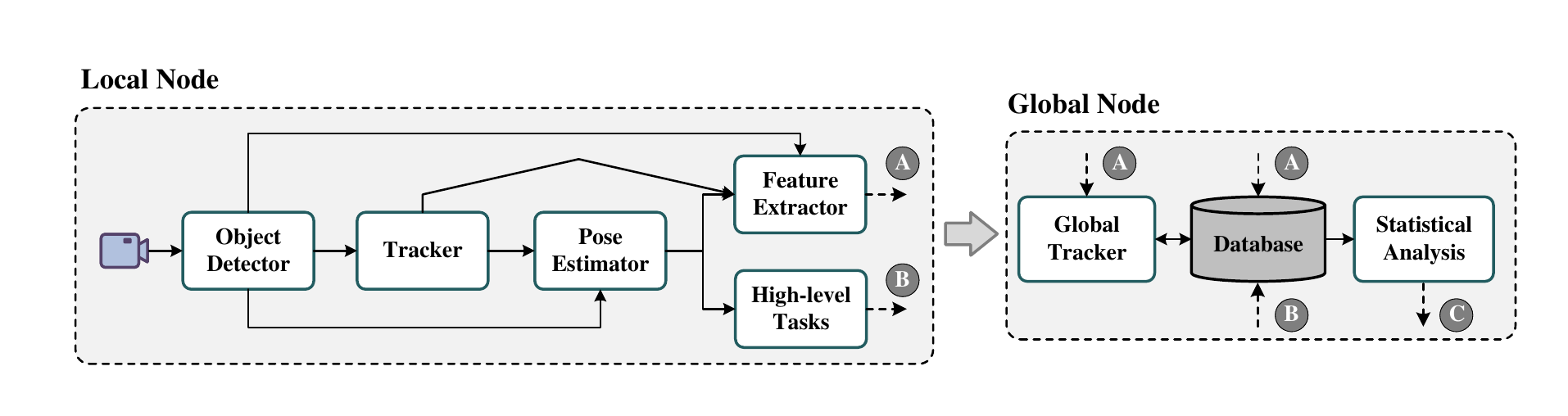}
                \caption{Data flow of the edge node. Solid arrows and dotted arrows represent intera-node and inter-node communication. $C$ is transmitted to the cloud node for further processing.}
                \label{fig:edge}
\end{figure*}

The overall data flow of the edge node can be seen in Figure \ref{fig:edge}. We adopt the Ancilia framework \cite{pazho2023ancilia} for the edge node in our system. The edge node comprises two sub-nodes: the Local Node and the Global Node. An object detector on the Local Node is utilized to identify and localize objects, such as pedestrians, within a streaming video frame. This produces the bounding boxes necessary for the subsequent steps. These bounding boxes are then filtered to exclude non-pedestrian instances and are fed into a re-identification (ID) algorithm, which assigns unique identifiers (local IDs) to each individual. The bounding boxes also serve as input for a pose estimator, which extracts pose on the human body. These poses contain sufficient information to aid high-level tasks, such as anomaly detection and action recognition, to achieve their goals without revealing identifiable information or demographics. Based on the quality of the extracted poses and bounding boxes, such as the confidence of the pose and the level of occlusion, the best representation of each individual is selected to extract a feature map. All this information, along with the output of any high-level tasks, is transferred to the Global Node for further processing.

The global node, which serves as a central hub for multiple local nodes, collects and stores data obtained from various sources. The local nodes, connected to the global node, transmit sets of information to it, which are then recorded and stored in a centralized database. This information is used to identify individuals globally, enabling individuals to be identified across multiple cameras and locations. Additionally, the global node is responsible for statistically analyzing the data stored in the centralized database. This analysis includes utilizing high-level information derived from the raw data collected by the local nodes. It is important to note that all information stored within the centralized database is anonymous, ensuring the protection of individuals' privacy. 

\subsection{Statistical Data Analysis}\label{Analysis}
The outputs of the global node are stored in a database on the local node. Table \ref{tab1} represents the type of stored data and their description in the database. As we can see, we do not store any images or any PII. Global IDs are the unique ids assigned to each detected object across multiple cameras. Global re-identification is a very crucial feature of the system. Global re-identification enables us to track individuals' gender, sex, age, and ethnicity neutrally across multiple cameras. The record time represents the actual time of the recorded videos. This data helps us analyze individuals' behaviors and detection across time. Camera-ID illustrates the camera that recorded each frame which can be used in the geospatial analysis. Finally, the bbox-tlwh includes the X and Y coordinates of the top left corner of each detected object's bounding box and the width and height of the bounding boxes. In fact, instead of storing the X and Y coordinates of all four corners of the bounding box, we store the coordinates of one point, and by providing the data of the width and height of the bounding box, the coordinates of the rest of the corners can be calculated. The Anomaly-scores show the score of detected anomalies in each frame ranging from 0 to 100, with 0 representing no anomalies and 100 showing an absolute anomaly. We set a threshold to define the anomalies. Actions and objects are categorical variables that show the type of actions and objects detected in the scene.  

\begin{table}[h]
\begin{center}
\caption{The description of the generated data in the edge node}\label{tab1}%
\resizebox{\linewidth}{!}{
\begin{tabular}{@{}llll@{}}
\toprule
\textbf{Generated Data}   & \textbf{Description} \\
\midrule
Global-ID    & The unique IDs assigned to each person across multiple cameras   \\
Record-time    & The record time of frames   \\
Camera-ID    & The unique ID of the camera that recorded the frame  \\
bbox-tlwh   & The XY coordinates of the top left point, width, and height of the bbox  \\
Anomaly-scores    & The scores of detected anomalies in each frame   \\
Actions   & The type of actions recognized in each frame   \\
Objects    & The type of objects identified in each frame   \\
\botrule
\end{tabular}}
\end{center}
\end{table}

\begin{figure*}[h]
        \centering
               \includegraphics[width=1\linewidth, trim= 5 5 5 5,clip]{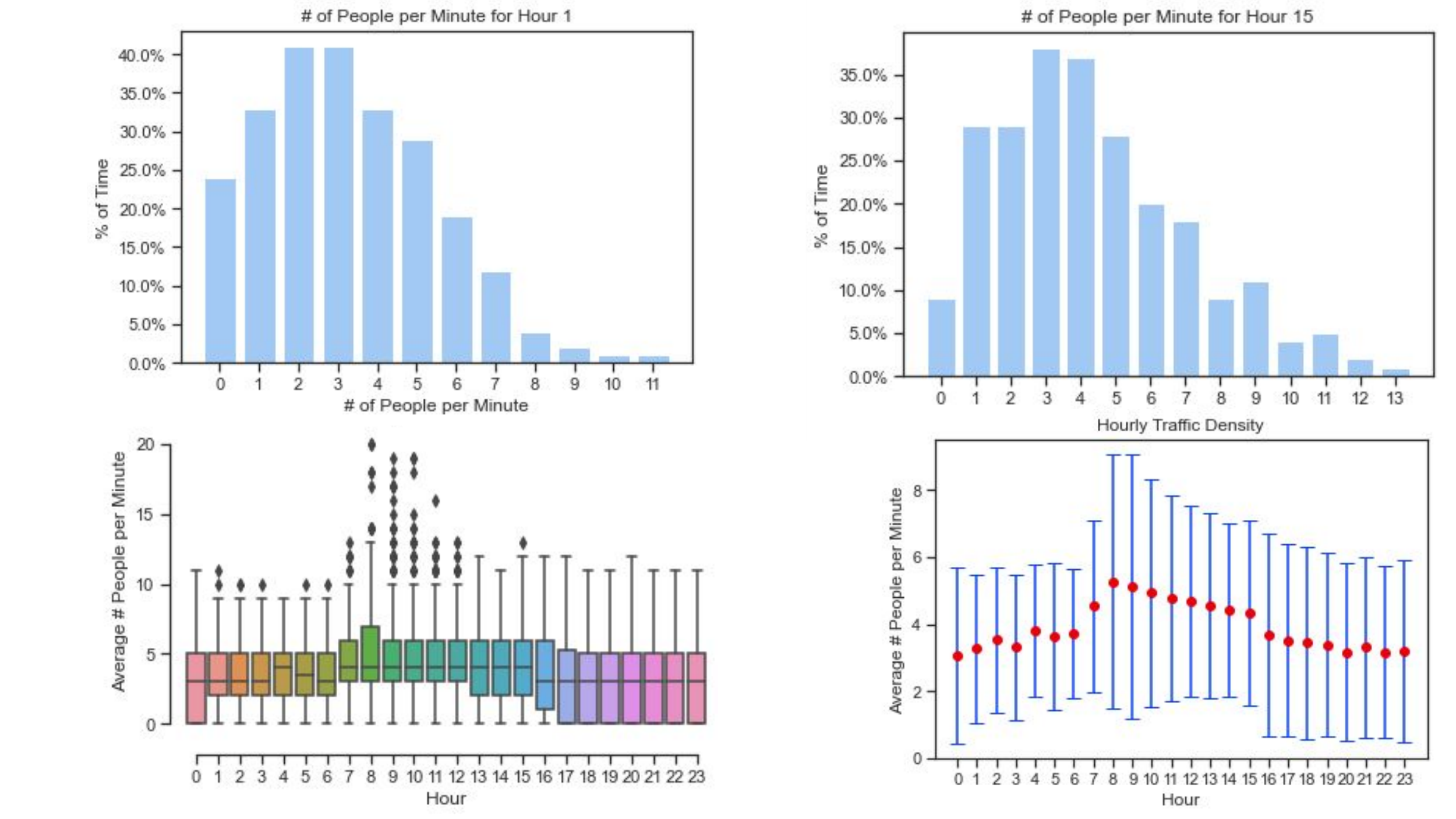}
                \caption{Examples of statistical analysis. Extracting distribution of human traffic and box plots that are used to identify the statistical anomalies and calculate occupancy indicator. }
                \label{fig:statistics}
\end{figure*}

These data can not be delivered to the end user because they need to reflect valuable insights. Secondly, they might be reversible, which increases the risk of privacy violations. Therefore, we analyzed these data on the edge node and sent the statistical analysis results to the cloud node. We use some statistical models to extract valuable information. First, by filtering the record time and counting the unique global IDs associated with each time, the number of people across each camera at across time is calculated. By extracting the distribution of the data over time, we can calculate the mean and standard deviation of the tracked people over time, enabling us to define the "statistical anomaly" metric. This metric shows an unexpected number of people respecting the historical data at each location. We also generate the plot box of the number of tracked people in each hour of the day by calculating the minimum, 25th percentile, 75th percentile, and maximum number of the detected people in each hour. Using these plot boxes, we generate the "Occupancy Indicator." This indicator compares the number of tracked individuals at any time to the historical data. It shows the user if any location is currently less than normal, normal, or over-occupied. Figure \ref{fig:statistics} represents some examples of distributions and box plots. Using the bbox-tlwh data, we can represent the Bird's eye view (BEV) of how people occupy each location. We use a transformation matrix to transmit the 2D coordinates to the BEV. Combining the BEVs over time enables us to generate the heat map that shows how people occupy each location over time. We also notify users in case of statistical anomalies. To inform the users of this type of anomaly, we calculate the mean and standard deviation of the tracked pedestrians over time by extracting the distribution of this data. Any new data is compared to the historical mean and standard deviation. If the new data is greater than the sum of the mean and two times of standard deviation, we call it a statistical anomaly and will notify the end user. The search feature of this application enables the end users to access historical data. They can search for the total number of people, the average number of people, and the maximum and the minimum number of people between any two-time points. These results are pushed to the cloud node to be stored and used for the smartphone application. 

\subsection{Cloud Node}\label{subsec5}
The cloud node receives analyzed data from the edge node. Cloud-native services provide robust data storage and management, user management solutions, and API generator services considering the scalability of the system\cite{dahunsi2021commercial}. Based on our objective, which is delivering only the necessary data to the end-users, we are using different Amazon Web Services (AWS) services to enable our end-to-end system to fulfill its goals. Since several cameras are in each location and there is a need to push real-time data to the cloud, a gateway is necessary to trigger AWS Lambda functions. This gateway executes specific actions such as notifying the end users, pushing data efficiently to the specified tables, and tracking messages. AWS Internet of Things (IoT) gateway enables us to configure how a message can interact with different services by defining necessary rules. Each rule consists of a Structured Query Language (SQL) SELECT statement that extracts data from incoming Message Queuing Telemetry Transport (MQTT) messages, a topic filter, and a rule action. Once a message is sent to the cloud, based on the topic and rule action, the gateway will push the data to the specified service using the MQTT protocol. We also need to ensure that users are notified of emergencies via their devices. A push notification service on the cloud provider is used in this regard. Therefore, necessary topics and messages are created on the gateway to enable the service to communicate with the rule-based message service. In the analysis section on the global node, emergency cases, such as detecting anomalous behavior, are distinguished and are pushed to the rule-based gateway as a specific topic. The gateway communicates with a rule-based message service to publish such messages on user devices.

We use a low-latency database to store data on the cloud to ensure the system can send real-time data. This low-latency database enables application developers to query the stored data with the key-value attribute\cite{dineva2021design}. We are using two types of tables to store data; tables that store the number of tracked objects across each camera and tables that store the analytical results associated with each camera over time. These tables are differentiated through the key-value attribute. We are using timestamps and camera ids as the key values. This set of key-value enables the client to access all data easily. As a result, on the user device, the users can search the database for desired statistics over time and in different locations. An application development service is required for the developer to generate the necessary Application programming Interfaces (API) for the smartphone application. This service creates a GraphQL schema to import data from existing tables. To avoid over-fetching data from the cloud to mobile and improve performance by retrieving data from the cloud efficiently, we used GraphQL.This two-way communication service uses the Hypertext Transfer Protocol Secure (HTTPS) protocol to send data to the end user's device. It enables end users to connect to the database and search for the desired statistics. User management and authentication are other aspects of the mobile application that cloud services handle. However, the details of such implementations go beyond the scope of this paper.
\subsection{Smartphone Application}\label{subsec6}
The final aim is to securely deliver the analyzed data generated on the edge node and statistical analysis section to the end user. We are developing a smartphone application to achieve this goal. This application consists of two main functions; delivering the analyzed data to the end user, and enabling the end user to search in the database. By using this smartphone application, end users can receive data such as real-time number of people at each location, real-time occupancy indicator using the historical data at each location, real-time bird's eye view of pedestrians across each camera, occupancy pattern of each location by generating heat map, type, time, and the number of anomalous behaviors, and cumulative data of the total and the average number of detected objects associated with each location over time. The end users will receive notifications on their devices if any anomalous behavior is detected. 
 
\textbf{Authentication and Authorization } are the most crucial elements in any end-to-end system. Users can register and log in to the mobile application with minimum details like first name, last name, email address, mobile number, and password. Users are verified using the temporary passcode sent to the email address before logging in to the home page. Users can only log in to the home page after validating the email address. The back end of the mobile application has been built using a cloud service that supports essential features like authentication, data modeling, API creation, and storage. All the user data is encrypted and stored on the cloud, providing maximum security and privacy. 
\begin{figure*}[h]
        \centering
               \includegraphics[width=0.8\linewidth]{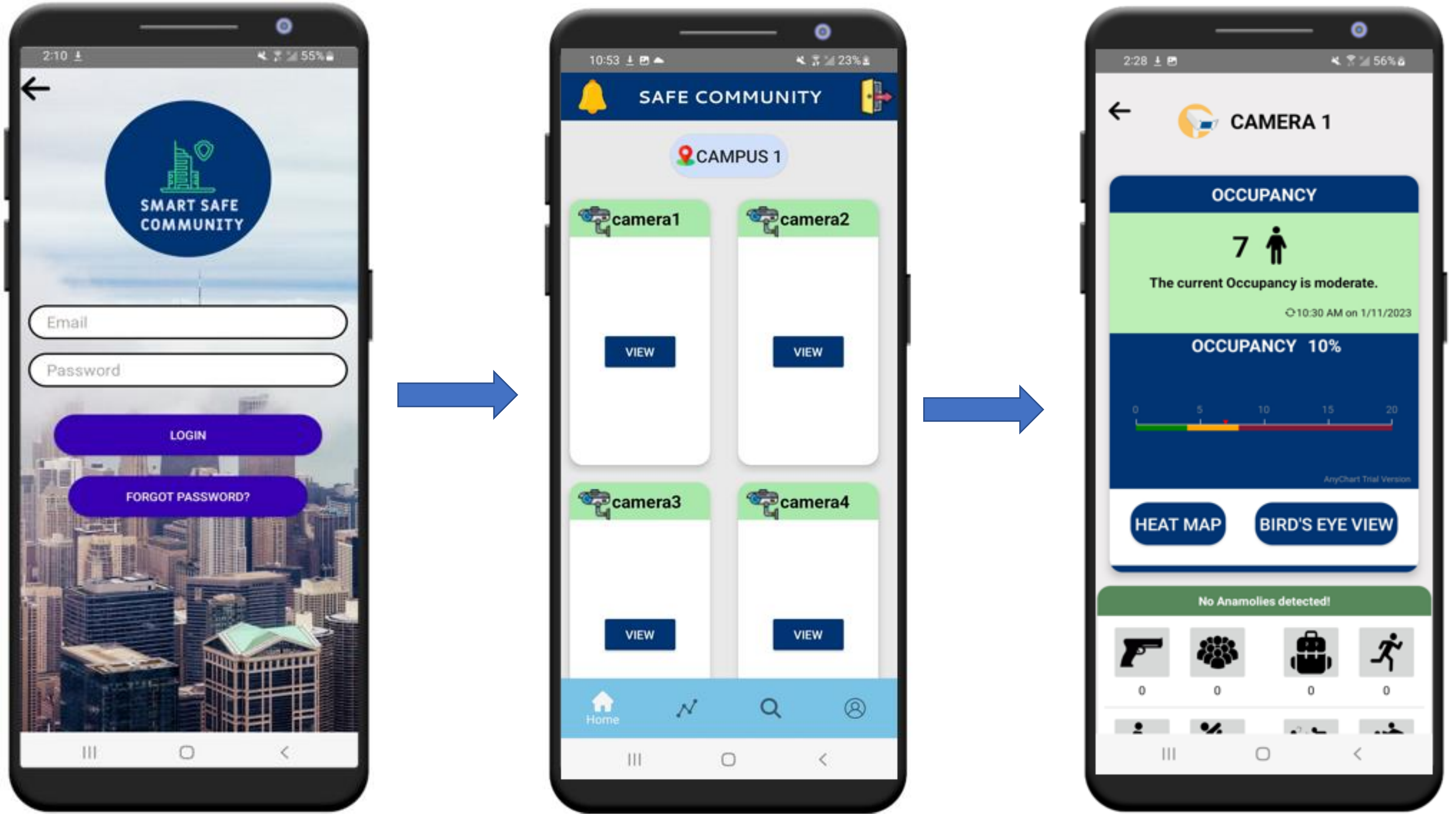}
                \caption{Image 1: The login interface of the mobile application. Image 2: The application's home screen displays all the cameras available at Campus 1. Image 3: The camera screen after selecting Camera 1 from the home screen.}
                \label{fig5}
\end{figure*}

\textbf{Home Screen} is the initial screen that a user sees after logging in. Considering design heuristics for the best user experience, this app screen was designed to navigate through different key features of the application swiftly. Users can navigate the dashboard, search, and profile screens on the bottom using the navigation bar. The toolbar on the top is designed with icons to log out and check notifications. This application is designed to support multiple cameras in different locations. users can tap on the location button and select the desired location. On the home screen, a list of all the cameras at that location is shown using card view. It contains the camera name and a green or red border to indicate whether the camera is live. On selecting a camera card view, it displays the camera screen.

\textbf{Camera Screen } contains all the detailed information about the camera. Under the Camera name, the first thing the user notices is the number of people identified at the location. Important information about the current situation at the location is displayed along with the timestamp when it was identified. This allows the users to track the current situation of the location in real-time without having any live feed or storing personal data of the people. The scale of the occupancy indicator changes with time, comparing the current number of people with the average of historical data at that given time. A pop-up displays a heat map and a bird's-eye view upon tapping the buttons below the occupancy indicator. Figure \ref{fig5} shown screenshots of the smartphone application.

\begin{figure*}[h]
        \centering
               \includegraphics[width=0.90\linewidth]{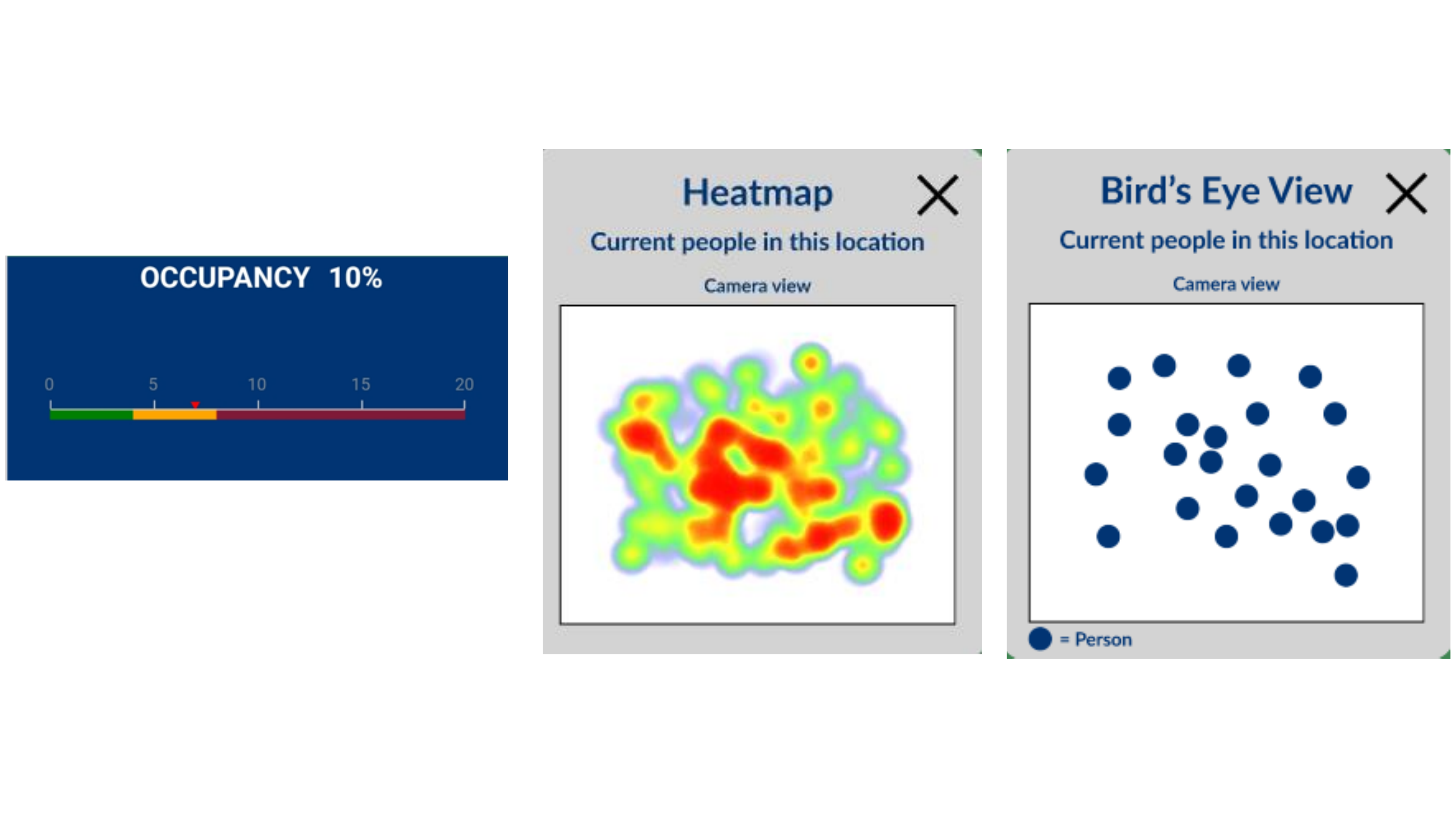}
                \caption{Image 1: Occupancy indicator showing current occupancy with respect to previous data. Image 2: Heat map from the camera view. Image 3: Bird's Eye view of the camera. }
                \label{fig6}
\end{figure*}

\textbf{Occupancy indicator} depicts the level of occupancy at the location based on historical data. A sample of this indicator is shown in \ref{fig6}. This gives the user insights into how occupied the space is at one glance. The number of people at every moment is stored on the database in the cloud from the AI pipeline. Over a period, this data helps to analyze the occupancy of every moment. This can be crucial information in predicting occupancy. The application of this indicator is enormous, as it can be used to schedule and allocate resources based on occupancy in educational institutions, hospitals, restaurants, and many other commercial and non-commercial domains.

\textbf{Heat map} is included in this screen to visualize and depict people's motion and behavior using different color codes. In figure \ref{fig6} an example of the generated heat map is presented. This offers the user a clear visual indication of the activity of the people across the camera’s view. This process aims to protect public privacy without compromising any features and take full advantage of the data obtained from the camera. The heat map is a 2-dimensional view of the camera depicted in different intensities of colors based on density. The heat map's X and Y coordinate values are stored on a cloud database service, then retrieved and displayed on the mobile app. The occupancy indicator lets the user understand how many people are there at a given time. Still, it cannot give any information about which area of the location people are spending their time. The heat map is the solution that aids in analyzing the data on where individuals spend their time. With this wide range of applications, for instance, in a clothing or grocery store using heat maps, we can analyze public time-spending patterns and understand which products or aisles the users prefer to spend more time on. In the same way, heat maps can be used in any field to provide solutions that a camera can provide and protect public privacy.

\textbf{Bird's-eye view} is the representation of a camera’s view transformed into a top-angle view and represented on an X and Y axis s can be seen in figure \ref{fig6}. This gives a great insight into how people are distributed at the location and how far each person is from others. The data obtained from the perspective transformation is stored on the cloud at every moment and retrieved and displayed on mobile applications. The heat map provides the perspective and an angle at which the camera is fixed, whereas the bird’s eye view is the powerful feature that transforms the camera’s angle to a top view. Almost everything a camera retains, video or images, can be accomplished without storing it using an occupancy indicator, heat map, and the bird’s eye view. This is achieved by maintaining public privacy.

\begin{figure*}[h]
        \centering
               \includegraphics[width=0.80\linewidth]{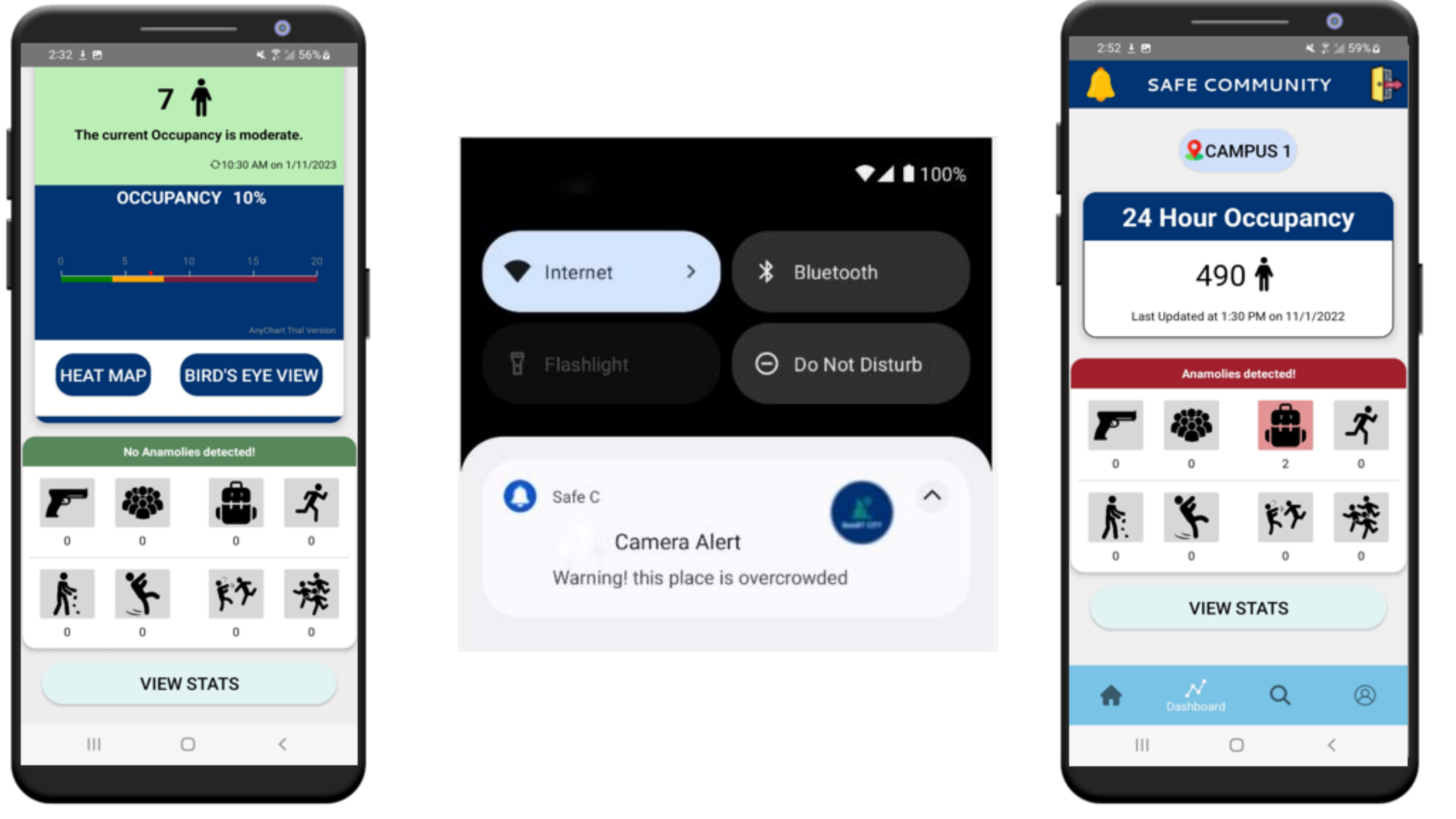}
                \caption{Image 1: Anomalies detected by Camera 1 Image 2: Push notification received on the app when Anomalies are detected. Image 3: Dashboard Screen of the app. Bag left behind anomaly detected.}
                \label{fig7}
\end{figure*}

\textbf{Anomalies} are the set of abnormal behaviors or actions that are identified at the camera location. Detecting anomalous behaviors or actions through surveillance camera footage is crucial in various environments.  The definition of anomalous behavior depends on the specific context and may vary across different settings. To effectively detect and respond to anomalous behavior, it is important to have the capability to derive multiple anomalies based on the requirements of the given environment. As can be seen in \ref{fig7} through the implementation of various detection algorithms and techniques, the system can identify and notify users of potential anomalies, such as the presence of firearms, mass gatherings, abandoned objects, and acts of violence. The user is presented with real-time updates on the identified anomalies and can access additional information through further examination. The stats button below the anomalies section displays the screen with the statistical analysis of camera 1 based on the number of people, timestamps,s and anomalies detected during the last 24 hours. 

\textbf{Notifications} are crucial for applications like this. It alerts the user when an anomaly is detected by notifying them with the push notification service. Users can be warned and prevent any unwanted things from happening. A push notification contains a title and a message. Users can also be notified using text messages and email if offline. Figure \ref{fig7} shows a screenshot of a push notification.

\textbf{Dashboard Screen} helps the users track the data of all the cameras that are "live" at each location. As shown in \ref{fig7} users can select the desired location and get the number of people identified by all the cameras, along with a timestamp. All the anomalies at each location and insightful graphs are displayed on this screen based on the last 24 hours of data.

\begin{figure*}[h]
        \centering
               \includegraphics[width=0.8\linewidth]{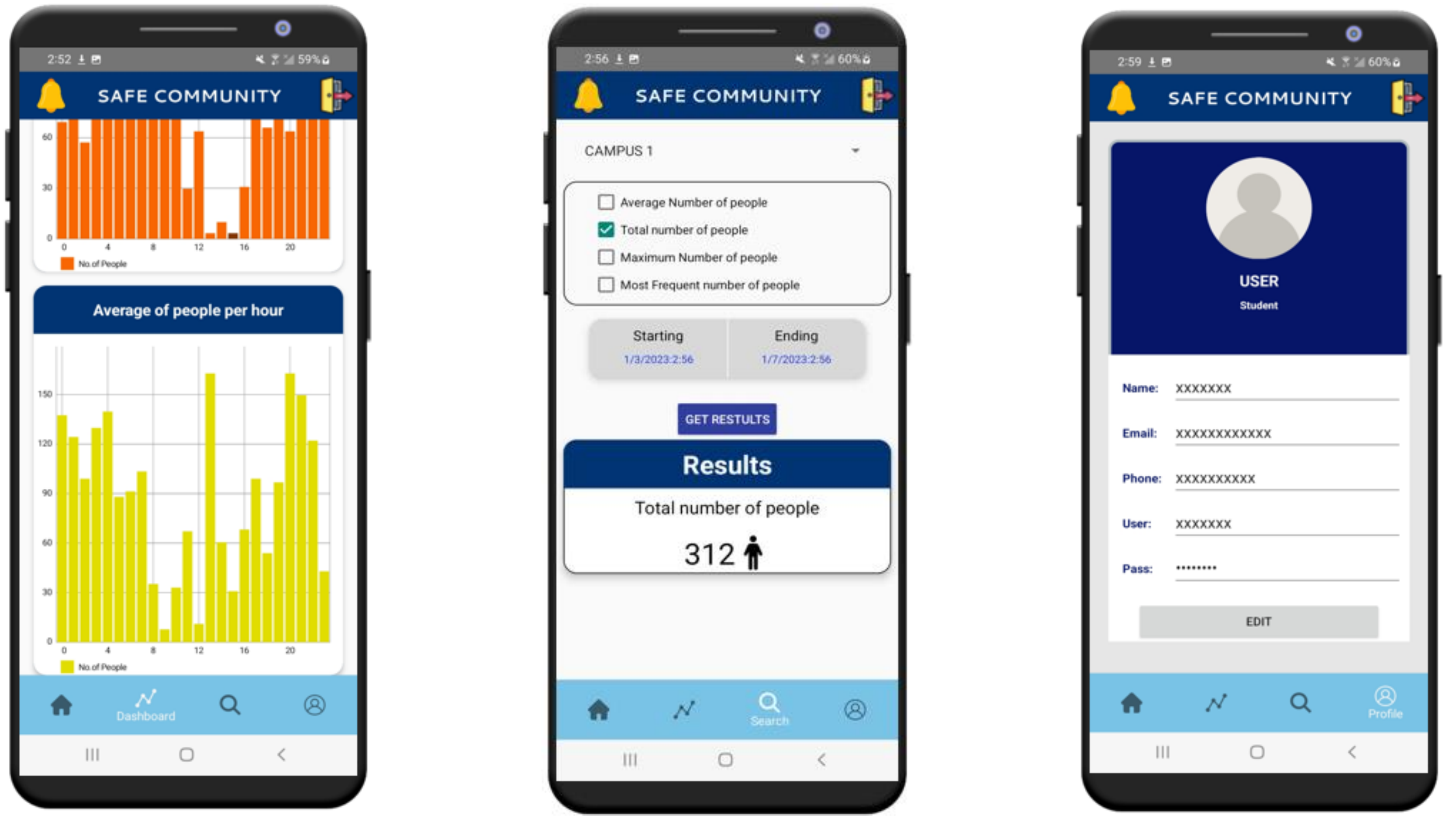}
                \caption{Image 1: Graph displaying the cumulative number of people for the last 24 hours. Image 2: Search screen to query the data stored on the cloud using different parameters. Image 3: Profile screen of the app displaying user data.}
                \label{fig8}
\end{figure*}

\textbf{Statistics  } that provide insightful information about the location and people's behavior where the camera is installed are processed on the cloud and displayed on the app. As shown in figure \ref{fig8} these stats are obtained based on the anomalies, the number of people assuring the privacy of the people. 

\textbf{Search Screen} helps users to select the location, date, and time, as well as the average, total, maximum, and most frequent number of people detected at the location. This provides flexibility to the user to query on desired date and time.

\textbf{ Profile Screen} displays the information of the user, and this can be updated at any time.

\section{Evaluation}\label{sec16}

\subsection{Quantitative Evaluation}\label{subsec7}

As discussed in Sec. \ref{edgenode}, on the edge node, following the footprints of Ancilia \cite{pazho2023ancilia}, YOLOv5 \cite{YOLOv5}, ByteTrack \cite{bytetrack}, HRNet \cite{hrnet}. Osnet \cite{osnet} are utilized for object detection, tracking, 2D pose extraction, and feature map representation production. Their performance and accuracy details can be seen in their corresponding references. 

To adopt all of these algorithms, it is necessary to have a powerful computing infrastructure. In our research, we utilized 2$\times$ 32 Cores 2.6 Ghz EPYC 7513 processors and 4$\times$ Nvidia V100 Graphics processing units (GPU). As shown in table \ref{edge_performance}, different scenes with different crowd densities result in different throughput and latency values. Normal (70 detections per second), heavy (216 detections per second), and extreme (744 detections per second) crowd densities result in a throughput of 52.94, 40.16, and 17.80 frames per second, respectively.
On top of that, for normal, heavy, and extreme crowd densities, the latency values are 5.39, 15.66, and 36.04 seconds respectively\cite{pazho2023ancilia}. It is important to note that the performance of the edge node is real-time, as it can process and analyze the frames as they are captured. This is a crucial aspect for many applications such as surveillance, autonomous navigation, and robotics, where real-time processing of visual data is required.

\begin{table}[]
\centering
\caption{Edge node performance evaluation.}
\label{edge_performance}
\begin{tabular}{lccc}
\toprule
                                                                                         & \textbf{\begin{tabular}[c]{@{}c@{}}Normal\\ Video Stream\end{tabular}} & \textbf{\begin{tabular}[c]{@{}c@{}}Heavy\\ Video Stream\end{tabular}} & \textbf{\begin{tabular}[c]{@{}c@{}}Extreme\\ Video Stream\end{tabular}} \\ \midrule
\textbf{\begin{tabular}[c]{@{}l@{}}Crowd Density\\ (Detections per Second)\end{tabular}} & 70                                                                     & 216                                                                   & 744                                                                     \\ \midrule
\textbf{\begin{tabular}[c]{@{}l@{}}Throughput\\ (Frames per Second)\end{tabular}}        & 52.94                                                                  & 40.16                                                                 & 17.80                                                                   \\ \midrule
\textbf{\begin{tabular}[c]{@{}l@{}}Latency\\ (Seconds)\end{tabular}}                     & 5.39                                                                   & 15.66                                                                 & 36.04                                                                  \\ \botrule
\end{tabular}%
\end{table}

As discussed, anomaly detection is our system's essential high-level machine-learning task. Anomaly detection in machine learning is a classification problem. In this task, a score is assigned to each frame, and based on the domain, a threshold is set to determine if a frame includes anomalous behavior. This threshold is domain-specific and could be obtained through expert panel discussions. Area Under the Curve (AUC) is the metric used to evaluate the performance of any multi-class classification problems, including anomaly detection tasks. AUC represents the models' ability to distinguish between different classes. The higher AUC shows that model is better at classifying a normal frame as normal and an anomalous scene as an anomaly\cite{narkhede2018understanding}. 


\begin{table}[h]
\centering
\caption{Comparison between pose-based and pixel-based high-level tasks. Anomaly Detection is evaluated on the ShanghaiTech dataset using the metric AUC-ROC. Action Recognition is evaluated on NTU60 X-Sub dataset and the reported number is Accuracy Percentile.}
\label{tab:my-table}

\begin{tabular}{@{}llll@{}}
\toprule
\multirow{4}{*}{Anomaly} & \multirow{2}{*}{Pose}  & GEPC \cite{markovitz2020graph}      & 73.72 \\ \cmidrule(l){3-4} 
                         &                        & MPED-RNN \cite{morais2019learning}  & 70.23 \\ \cmidrule(l){2-4} 
                         & \multirow{2}{*}{Pixel} & S3R \cite{wu2022self}       & 97.48 \\ \cmidrule(l){3-4} 
                         &                        & RFTM \cite{tian2021weakly}      & 97.21 \\ \midrule
\multirow{4}{*}{Action}  & \multirow{2}{*}{Pose}  & PoseConv3D \cite{duan2022revisiting}& 92.76 \\ \cmidrule(l){3-4} 
                         &                        & CTR-GCN \cite{chen2021channel}   & 83.07 \\ \cmidrule(l){2-4} 
                         & \multirow{2}{*}{Pixel} & MMNet \cite{bruce2022mmnet}     & 96.0  \\ \cmidrule(l){3-4} 
                         &                        &       VPN \cite{das2020vpn}    &     95.5  \\ \bottomrule
\end{tabular}%

\end{table}

AUC-ROC stands for "Area Under the Receiver Operating Characteristic Curve." It is a metric used to evaluate the performance of a binary classifier. The ROC curve plots the true positive rate (TPR) against the false positive rate (FPR) at various threshold settings. The AUC represents the classifier's overall performance by measuring the area under this curve. AUC-ROC is a value between 0 and 1, where a value of 1 represents a perfect classifier, and a value of 0.5 represents a classifier no better than random guessing \cite{olson2008advanced}.

It is worthwhile to compare the results of an anomaly detection algorithm without privacy consideration against our model to see the cost of designing a privacy perseverance system. Generally, SVS systems use two types of algorithms to conduct deep learning tasks: Pixel-based and pose-based algorithms. To fulfill the goals of the system, we are using pose-based algorithms. Further explanations of the reasons for preferring pose-based to pixel-based algorithms are provided in the Qualitative Evaluation section. Wu et al. reported the reported AUC as 97.48 in their proposed self-supervised video anomaly detection trained on ShahghaiTech dataset\cite{wu2022self}.
On the other hand, we used GEPC \cite{markovitz2020graph}, and MPED-RNN \cite{morais2019learning}, on the shahgahiTech dataset to test our models. Our results show 73.72 AUC based on GEPC and 70.23 based on MPED-RNN \cite{pazho2023ancilia}. As shown in table \ref{tab:my-table}, the AUC-ROC results for both action and anomaly detection tasks dropped by using pose-based algorithms. The reason is that pixel-based approaches are trained based on more data points. 

In addition to the vision pipeline latency, the cloud node's latency is also essential. As discussed, we are using different cloud-native services to deliver the data to the end users. Among the services we use, DynamoDB, Amplify, and AppSync are likely to cause latency in the system. It is shown in table \ref{tab5} that according to the last two months' records, CloudWatch (Amazon AWS service to monitor clouds' activities) reports on average 0.06 seconds latency for Amplify, 96.3 milliseconds latency for Appsync, and 41.4 milliseconds latency for DynamoDB. However, the CloudWatch results show that Appsync and Amplify experience latency in a few cases and are not continuous over time. Increasing the number of users at specific times is the source of this latency which should be addressed in future versions of the application. Therefore, the overall latency of the end-to-end system in an extreme crowd scene (36.04 seconds) with multiple users (0.06 seconds) will be 36.1 seconds on average, which should be improved. 

\begin{table}[h]
\begin{center}
\caption{Cloud-native services latency}\label{tab5}%
\begin{tabular}{@{}llll@{}}
\toprule
\textbf{Service}   & \textbf{Average Latency} & \\
\midrule
DynamoDB Table    &  41.4 ms   \\
AppSync    & 96.3 ms   \\
Amplify    & 0.06 s  \\
\botrule
\end{tabular}
\end{center}
\end{table}

\subsection{Qualitative Evaluation}\label{subsec8}
We discussed this design's algorithms, services, and data flow in the proposed system section. We qualitatively evaluate the system using the four levels of privacy perseverance discussed in the Privacy Perseverance System Features section. All AI-based algorithms are pose-based algorithms in terms of algorithms. We avoid using algorithms that use identifiable information, such as pixel-based algorithms, in this design. The system's most critical components, such as anomaly detection, action recognition, and global re-identification, use skeleton and abstract feature representation. Because of this method of approaching the algorithm, neither the inputs nor the outputs are identifiable information. Furthermore, the outputs are race, gender, age, and ethnicity agnostic. This aspect of the system addresses discrimination as a fundamental ethical challenge in the domain of public safety\cite{nissenbaum2004privacy}.

Data transmission is a critical component of system design in the SVS context \cite{nissenbaum2004privacy, hartzog2018privacyos}. Cameras in SVS systems can capture images of people. These images can be used directly by the data collector for processing or sold to a third party by the data collector. Whether the data is used by the data collector or sold to a third party, the system's designer should consider the essential security practices to prevent image and data leaks. Even though our system does not rely on identifiable information, we must ensure that the information cannot be transferred to an unauthenticated party. To address this issue, we are using a local server protected by multiple firewalls and can only be accessed by verified users. The SVS pipeline and de-identified information database are hosted on a local server, as discussed earlier in this paper and Figure \ref{fig:E2E}. We also analyze the data on the local server before sending it to a cloud-based server.

All privacy protection legislation focuses on data retention and irreversibility, meaning outputs should not be identifiable or reversible when selecting appropriate models for machine learning tasks. To address this issue, we are taking two approaches. First and foremost, we do not employ facial recognition technologies. Pose-based models are used in all pipeline sections described in the Proposed System section. Using pose-based models ensures that our system does not use identifiable data like images.
Furthermore, because we are not storing the image frames in any part of the system, no one has access to the images captured by the cameras. Even though we are training the models with abstract feature representations, the global re-identification model may have a reversibility issue. The global re-identification model uses bounding box features to identify objects across multiple cameras\cite{ye2021deep}. As a result, the global re-ID model should be able to store these features and, once a person is detected, cross-check the new person's features with the stored features to determine whether this is a new person. These features can be identified from the model's perspective by reversing back. In theory, if someone has access to each person's abstract feature representations and the neural network model's weights, he or she can decode the model and recover the main image to an acceptable resolution level \cite{radenovic2018fine}. Our second approach is to solve the reversibility issue. For global re-ID, we propose to use online learning. This method updates the model weights every 30 minutes, and previous weights are automatically destroyed. As a result, even if someone has access to the extracted features, there is no way to restore the images.

The type of stored data is vital for a system to comply with privacy protection acts, according to these acts. As discussed earlier in this paper, we have considered not only considerations regarding stored data but also ethical considerations at the data processing level. According to our system features, we do not use pixel-based algorithms, as previously stated. Furthermore, we must store and transfer the actual videos or images. When these two are combined, the output data is de-identified. However, our local and global re-ID algorithms do not use facial recognition technology, ensuring that we do not use any personally identifiable information to identify individuals. Table \ref{tab2} summarizes our solutions to address the pre-mentioned privacy challenges. 

\begin{table}[h]
\begin{center}
\caption{The proposed system's solutions to address policy challenges}\label{tab2}%
\begin{tabular}{@{}llll@{}}
\toprule
\textbf{Metrics}   & \textbf{Solution}\\
\midrule
Algorithm  & Using Pose-Based Algorithms    \\
System     & Using Local Server   \\
Model     & Making Data Irreversible   \\
Data    & Not Using PII   \\
\botrule
\end{tabular}
\end{center}
\end{table}

 \subsection{Discussion}
 Our proposed end-to-end system design provides a road map for a more holistic approach to addressing ethical challenges in SVS system design. The SVS presents some unique privacy challenges. People's privacy can be violated when PII and facial recognition technology are used in processing. Actual video storage and transfer can increase the likelihood of privacy violations.
We argued that when designing an SVS system, privacy concerns could be addressed from four perspectives: algorithm, system, model, and data. Pose-based algorithms are used in the algorithm to ensure that no identifiable information is used. To increase the system's security, we run all pipeline algorithms on a local server. We also use online learning methods for local and global re-identification as a solution to reversible data. We do not store images or videos in terms of data.

We argued that the system is generally designed to address privacy concerns; however, privacy is only one of the ethical concerns addressed by the current system. Discrimination is currently a critical ethical issue in policing society, according to Miller(2017) \cite{miller2017ethical}. Because we do not use personally identifiable information or facial recognition technology, our system is racial, age, and gender-neutral, providing a fundamental baseline for removing biases in the policing and monitoring processes. 

Our end-to-end system is fully functional, as demonstrated in the Evaluation section. However, the results must still be cutting-edge and improved. Indeed, at this early stage, our focus is on two things: system functionality and addressing the ethical issue of privacy. As a result, we are sacrificing accuracy to respect privacy in the current configuration. The next step will be to optimize the system's algorithms and services on both the local and cloud servers. 

\begin{figure}[h]
        \centering
               \includegraphics[width=1\linewidth]{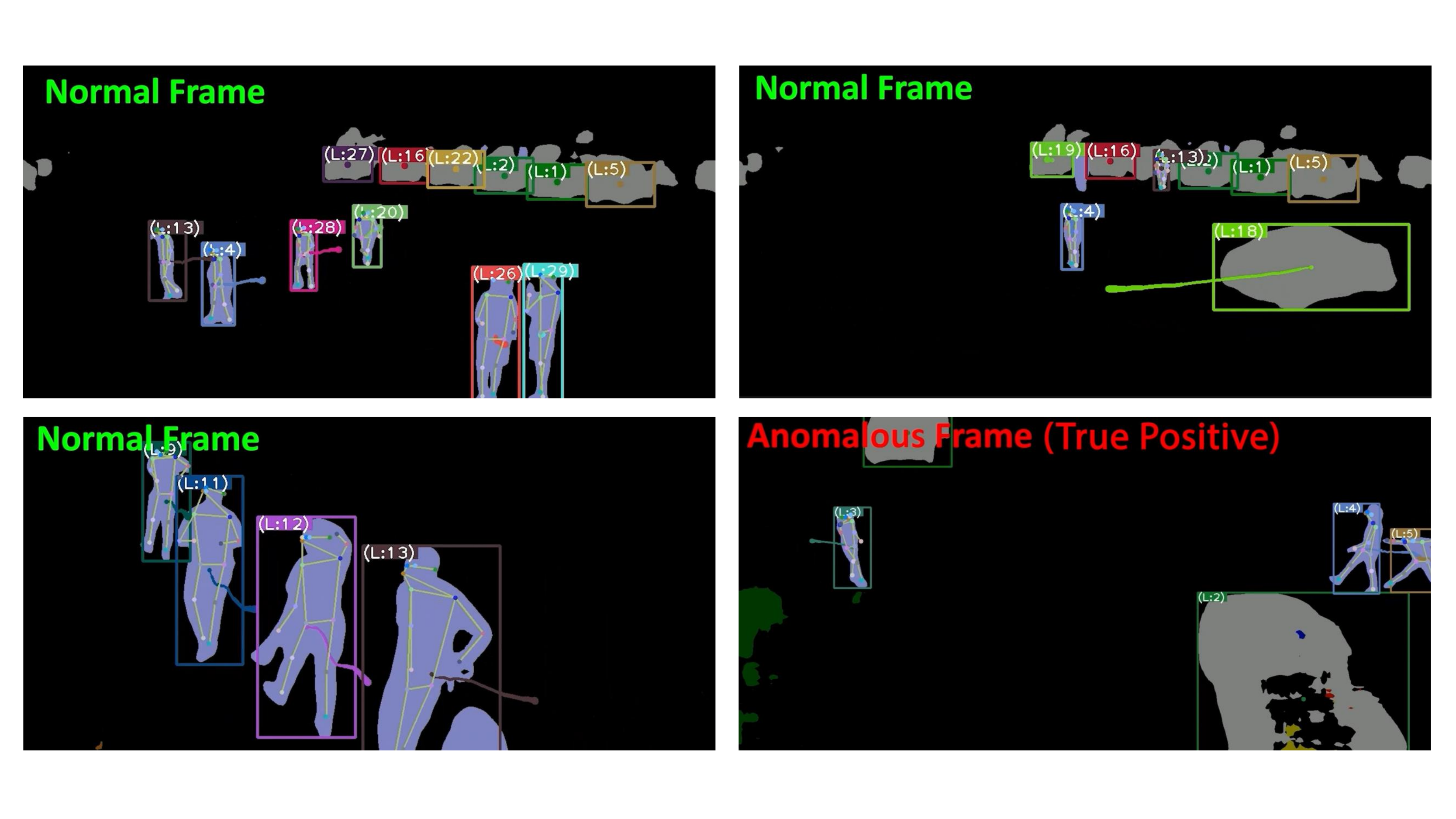}
                \caption{Examples of frame labeling. The green labeled frames are normal frames, while in the last frame, an anomaly is detected, and the system correctly distinguishes the anomaly.}
                \label{fig:frames}
\end{figure}

According to the definition, false negatives and positives are crucial metrics in evaluating the system. A lower AUC-ROC means that the classifier is not correctly putting each input in its corresponding category. This can be roughly translated to False Positives and False Negatives or the system's incorrect decisions. High false positives in an anomaly detection system mean the system labels normal scenes as anomalies, eventually losing trust in the technology. On the other hand, false negatives will cause liability problems. Because, in this case, the system can not detect anomalies. Figure \ref{fig:frames} represents examples of normal frames and a True Positive anomalous frame. Most classification algorithms are skewed to minimize false negatives in cost having a reasonable level of false positives\cite{viola2001fast}. However, given the role of anomaly detection in AI-enabled systems for building safer communities, the goal is to achieve the minimum possible level of false positives under the condition that false negatives are zero. 
This system can be used in various public domains to assist communities in improving public safety. It can also be used as a more efficient alternative to current passive surveillance systems by the public and private sectors. Examples of sectors that can benefit from this system include public parking lots, grocery store parking lots, university campuses, bus and train stations, city centers, and plazas. Although the system's primary goal is to ensure safety through a privacy preservation system, private and public sectors can benefit from the information provided. This information has the potential to provide insightful business solutions for the sector. 

Improving the system's overall functionality, improving accuracy, lowering latency, and optimizing bandwidth usage, Central processing unit (CPU), and GPU usage are all possible future works. Eliminating false negatives and conducting usability studies regarding the users' preferences in using such technology could be another research direction. More advanced statistical analysis is required in terms of data. Considering the currently available data, such as bird's eye view, heat map, action recognition, and anomaly detection, sociologists may find it very interesting to investigate various social issues concerning occupied spaces. For instance, if any individual or group actions are more likely to occur in a particular location. Another aspect that could be addressed is researching the factors that can lead to communities engaging with this system. Finally, because we only looked at privacy issues as ethical challenges in the context, there is still room to investigate the impact of other ethical issues, such as trust, in designing SVS systems to provide safety to society.  

\section{Conclusion}\label{sec16}
This article discussed the privacy challenges of designing smart video surveillance systems and provided an overview of regulations for addressing these challenges. Based on this discussion, we proposed an end-to-end privacy-preserving system at four levels: Video analytics algorithms, statistical analysis models, cloud-native services, and smartphone applications. We defined both quantitative and qualitative metrics for evaluating such a system. Our approach toward designing an AI-enabled SVS system shows how considering privacy concerns will affect different elements of such a system. Although the system performed acceptably in extreme video scenes with 17.8 FPS, the accuracy of the high-level tasks such as action detection and anomaly detection dropped in the cost of considering privacy, which should be addressed in the future. We selected pose-based algorithms to perform these tasks, and as a result, the results dropped from 97.48 to 73.72 in anomaly detection and 96 to 83.07 in the action detection task. On the other hand, the average latency of the end-to-end system was 36.1 seconds, which is noticeably high in the context. These results show that from the technical perspective, privacy can be addressed at the design level; however, it still needs to be improved to be used in real-world scenarios. 

\section{Abbreviations}
\begin{itemize}
  \item AI (Artificial Intelligence) 
  \item SVS (Smart Video Surveillance)
  \item FPS (Frame-Per-Second)
  \item  CCTV (Closed Circuit TVs)
  \item CAGR (Compound Annual Growth Rate)
  \item HIPAA (Health Insurance Portability and Accountability Act (Message Queuing Telemetry Transport) 
  \item  CPPA (California Consumer Privacy Act) 
  \item  ADPPA (American Data Privacy and Protection Act)
  \item  GDPR (General Data Protection Regulation)
  \item  PHI (Protected Health Information)
  \item  CMIA (Confidentiality of Medical Information Act)
  \item  CPRA (California Consumer Privacy Rights Act)
  \item  EU (European Union)
  \item  EEA (European Economic Area)
  \item  PII (Personally Identifiable Information)
  \item  SSN (Social Security Number)
  \item  ID (Identification)
  \item  BEV (Bird's Eye View)
  \item  AWS (Amazon Web Services)
  \item  IoT (Internet of Things)
  \item  SQL (Structured Query Language)
  \item  MQTT (Message Queuing Telemetry Transport)
  \item  API (Application Programming Interface)
  \item  HTTPS (Hypertext Transfer Protocol Secure)
  \item  GPU (Graphics Processing Units)
  \item  AUC (Area Under the Curve)
  \item  AUC-ROC (Area Under the Receiver Operating Characteristic Curve)
  \item  TPR (True Positive Rate)
  \item  FPR (False Positive Rate)
  \item  CPU (Central Processing Unit)
\end{itemize}
\section* {Acknowledgment}
This research is supported by the National Science Foundation (NSF) under Award No. 1831795.

\bibliography{sn-bibliography}

\end{document}